\documentclass[conference]{IEEEtran}

\DeclareMathAlphabet{\mathitbf}{OML}{cmm}{b}{it}

\usepackage{tikz}
\usetikzlibrary{matrix,arrows}
\usetikzlibrary{shapes.geometric, arrows}

\usepackage{amsmath,amssymb,mathrsfs,amsthm}
\usepackage[colorlinks=true,linkcolor=black,anchorcolor=black,citecolor=black,filecolor=black,menucolor=black,urlcolor=black]{hyperref}
\usepackage{graphicx}
\usepackage{psfrag}
\usepackage{subcaption}
\usepackage{cite}
\usepackage{multirow}
\usepackage[margin=1in]{geometry}
\usepackage[ruled,vlined]{algorithm2e}

\usepackage{epsfig}
\usepackage{amsfonts}
\usepackage{graphics}
\usepackage{color}
\usepackage{mathtools}
 
\epsfverbosetrue

\usepackage{xspace}
\usepackage{bbm}
\catcode`@=11
\chardef\mathlig@atcode\count255

\def\actively#1#2{\begingroup\uccode`\~=`#2\relax\uppercase{\endgroup#1~}}
\def\mathlig@gobble{\afterassignment\mathlig@next@cmd\let\mathlig@next= }
\def\mathlig@delim{\mathlig@delim}
\def\mathlig@defcs#1{\expandafter\def\csname#1\endcsname}
\def\mathlig@let@cs#1#2{\expandafter\let\expandafter#1\csname#2\endcsname}
\def\mathlig@appendcs#1#2{\expandafter\edef\csname#1\endcsname{\csname#1\endcsname#2}}

\def\mathlig#1#2{\mathlig@checklig#1\mathlig@end\mathlig@defcs{mathlig@back@#1}{#2}\ignorespaces}

\def\mathlig@checklig#1#2\mathlig@end{%
 \expandafter\ifx\csname mathlig@forw@#1\endcsname\relax
 \expandafter\mathchardef\csname mathlig@back@#1\endcsname=\mathcode`#1%
 \mathcode`#1"8000\actively\def#1{\csname mathlig@look@#1\endcsname}%
 \mathlig@dolig#1\mathlig@delim
\fi
\mathlig@checksuffix#1#2\mathlig@end
}

\def\mathlig@checksuffix#1#2\mathlig@end{%
\ifx\mathlig@delim#2\mathlig@delim\relax\else\mathlig@checksuffix@{#1}#2\mathlig@end\fi
}
\def\mathlig@checksuffix@#1#2#3\mathlig@end{%
\expandafter\ifx\csname mathlig@forw@#1#2\endcsname\relax\mathlig@dosuffix{#1}{#2}\fi
\mathlig@checksuffix{#1#2}#3\mathlig@end
}

\def\mathlig@dosuffix#1#2{%
\mathlig@appendcs{mathlig@toks@#1}{#2}%
\mathlig@dolig{#1}{#2}\mathlig@delim
}

\def\mathlig@dolig#1#2\mathlig@delim{%
 \mathlig@defcs{mathlig@look@#1#2}{%
 \mathlig@let@cs\mathlig@next{mathlig@forw@#1#2}\futurelet\mathlig@next@tok\mathlig@next}%
 \mathlig@defcs{mathlig@forw@#1#2}{%
  \mathlig@let@cs\mathlig@next{mathlig@back@#1#2}%
  \mathlig@let@cs\checker{mathlig@chck@#1#2}%
  \mathlig@let@cs\mathligtoks{mathlig@toks@#1#2}%
  \expandafter\ifx\expandafter\mathlig@delim\mathligtoks\mathlig@delim\relax\else
  \expandafter\checker\mathligtoks\mathlig@delim\fi
  \mathlig@next
 }%
 \mathlig@defcs{mathlig@toks@#1#2}{}%
 \mathlig@defcs{mathlig@chck@#1#2}##1##2\mathlig@delim{%
  \ifx\mathlig@next@tok##1%
   \mathlig@let@cs\mathlig@next@cmd{mathlig@look@#1#2##1}\let\mathlig@next\mathlig@gobble
  \fi 
  \ifx\mathlig@delim##2\mathlig@delim\relax\else
   \csname mathlig@chck@#1#2\endcsname##2\mathlig@delim
  \fi
 }%
 \ifx\mathlig@delim#2\mathlig@delim\else
  \mathlig@defcs{mathlig@back@#1#2}{\csname mathlig@back@#1\endcsname #2}%
 \fi
}%

\catcode`@\mathlig@atcode

\newcommand{\muspace}{\mspace{1mu}}

\DeclareRobustCommand{\scond}{\mathchoice{\muspace\vert\muspace}{\vert}{\vert}{\vert}}
\mathlig{|}{\scond}
\newcommand{\cond}{\mathchoice{\,\vert\,}{\mspace{2mu}\vert\mspace{2mu}}{\vert}{\vert}}

\DeclareRobustCommand{\discint}{\mathchoice{\mspace{-1.5mu}:\mspace{-1.5mu}}{\mspace{-1.5mu}:\mspace{-1.5mu}}{:}{:}}
\mathlig{::}{\discint}
\newcommand{\suchthat}{\mathchoice{\colon}{\colon}{:\mspace{1mu}}{:}}

\newcommand{\Kc}{\mathcal{K}}

\newcommand{\Cr}{\mathscr{C}}

\def\textiid{i.i.d.\@\xspace}
\newcommand\iid{\ifmmode\text{ i.i.d. } \else \textiid \fi}

\def\mathllap{\mathpalette\mathllapinternal}
\def\mathllapinternal#1#2{%
  \llap{$\mathsurround=0pt#1{#2}$}}

\def\clap#1{\hbox to 0pt{\hss#1\hss}}
\def\mathclap{\mathpalette\mathclapinternal}
\def\mathclapinternal#1#2{%
  \clap{$\mathsurround=0pt#1{#2}$}}

\let\oldstackrel\stackrel
\renewcommand{\stackrel}[2]{\oldstackrel{\mathclap{#1}}{#2}}

\renewcommand{\hbar}{h\mathllap{\overline{\vphantom{h}\hphantom{\rule{4.6pt}{0pt}}}\mspace{0.77mu}}}

\catcode`~=11 
\newcommand{\urltilde}{\kern -.06em\lower -.06em\hbox{~}\kern .02em}
\catcode`~=13 

\def\0{\bf{0}}

\newtheorem{theorem}{Theorem}
\newtheorem{lemma}{Lemma}
\newtheorem{proposition}{Proposition}

\theoremstyle{definition}
\newtheorem{definition}{Definition}

\newtheorem{corollary}{Corollary}

\newtheorem{remark}{Remark}

\begin{document}

\title{On Critical Index Coding Problems}
\author{
\authorblockN{Fatemeh Arbabjolfaei and Young-Han Kim}
\authorblockA{Department of Electrical and Computer Engineering\\
University of California, San Diego\\
Email: \{farbabjo, yhk\}@ucsd.edu
}
}
\date{}
\maketitle

\begin{abstract}
The question of under what condition some side information for index coding can be removed without 
affecting the capacity region is studied, which was originally posed by 
Tahmasbi, Shahrasbi, and Gohari.
To answer this question, the notion of
unicycle for the side information graph is introduced and it is shown
that any edge that belongs to a unicycle 
is critical, namely, it cannot be removed without reducing the capacity region.
Although this sufficient condition for criticality is not necessary in general, a partial
converse is established, which elucidates the connection between
the notion of unicycle and the maximal acylic induced subgraph outer bound on the
capacity region by Bar-Yossef, Birk, Jayram, and Kol.
\end{abstract}

\section{Introduction}

The index coding problem is a canonical problem in network information theory in which a server has a tuple of $n$ messages $x^n = (x_1, \ldots, x_n)$, $x_j \in \{0,1\}^{t_j}$, and is connected to $n$ receivers via a noiseless broadcast channel.
Receiver $j \in [1:n] := \{1,2, \ldots, n\}$ is interested in message $x_j$ and has a subset of other messages $x(A_j) := (x_i, i \in A_j), A_j \subseteq [1:n] \setminus \{j\}$ as side information.
Assuming that the server knows side information subsets, $A_1, \ldots, A_n$, the goal is to characterize the minimum number of transmissions the server needs to make such that each receiver can recover its desired message.

\begin{figure}[t]
\begin{center}
\small
\psfrag{1}[cb]{1}
\psfrag{2}[rc]{2}
\psfrag{3}[lc]{3}
\psfrag{4}{4}
\includegraphics[scale=0.45]{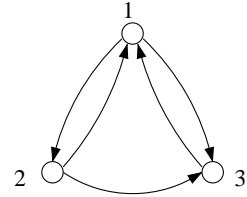}
\end{center}
\caption{The  graph representation for the index coding problem with $A_1 = \{2,3\}, A_2 = \{1\}$, and $A_3 = \{1,2\}$.}
\label{fig:3-message}
\end{figure}

Any instance of the index coding problem is fully determined by the side information subsets $A_1, \ldots, A_n$. 
An equivalent specification of the problem is the side information graph which is defined to be a directed graph with $n$ nodes.
Each node corresponds to a receiver and there is a directed edge $i \to j$ if and only if receiver $j$ knows message $i$ as side information (see Fig. \ref{fig:3-message}).
In this paper, we often refer to an index coding problem with its side information graph 
$G = (V,E)$ and write ``index coding problem $G$.''

A $(t_1, \ldots, t_n, r)$ index code is defined by
\begin{itemize}
\item an encoder $\phi: \prod_{i=1}^n \{0,1\}^{t_i} \to \{0,1\}^r$ that maps $n$-tuple of messages $x^n$
to an $r$-bit index and
\item $n$ decoders $\psi_j: \{0,1\}^r \times \prod_{k \in A_j} \{0,1\}^{t_k} \to \{0,1\}^{t_j}$ that maps the received index $\phi(x^n)$ and the side information $x(A_j)$ back to $x_j$ for $j \in [1::n]$.
\end{itemize}
Thus, for every $x^n \in \prod_{i=1}^n \{0,1\}^{t_i}$,
\[
\psi_j(\phi(x^n), x(A_j)) = x_j, \quad j \in [1::n].
\]
A rate tuple $(R_1,\ldots,R_n)$ is said to be \emph{achievable} for the index coding problem $G$
if there exists a $(t_1, \ldots, t_n, r)$ index code such that 
\[
R_j \leq \frac{t_j}{r}, \quad j \in [1:n].
\]
The \emph{capacity region} $\Cr$ 
of the index coding problem is defined as the closure of the set of achievable rate tuples.
The symmetric capacity of the index coding problem is also defined as
\[ 
C_\mathrm{sym} = \max \{R \suchthat (R, \ldots, R) \in \Cr\}.
\]

The index coding problem was introduced in 1998 by Birk and Kol \cite{Birk--Kol1998} in the context of satellite communication.
Since then, it has been studied by researchers in diverse areas
using algebraic \cite{Bar-Yossef--Birk--Jayram--Kol2011, Jafar2014, Maleki--Cadambe--Jafar2014}, graph theoretical \cite{Blasiak--Kleinberg--Lubetzky2013,Shanmugam--Dimakis--Langberg2013b}, and random coding \cite{Arbabjolfaei--Bandemer--Kim--Sasoglu--Wang2013} tools.
However, none of the proposed inner and outer bounds on the capacity region is tight in general, and
the problem of even approximating the capacity region within a factor of $O(n^{1-\epsilon})$ still remains open.

In \cite{Tahmasbi--Shahrasbi--Gohari2014b}, Tahmasbi, Shahrasbi, and Gohari asked a much
simpler question of how a single edge in the side information graph can affect the capacity region.
Their question is captured formally by the notion of criticality of an edge.

\begin{definition}
Given an index coding problem $G = (V,E)$,
an edge $e \in E$ is said to be critical if the 
removal of $e$ from $G$ strictly reduces the capacity region.
\end{definition}

\begin{definition}
The index coding problem $G = (V,E)$ is said to be critical if every $e \in E$ is critical.
\end{definition}

Thus, each critical graph (= index coding problem) cannot be made ``simpler'' into another one of
the same capacity region. 
In the following, we recall two necessary conditions for criticality.

\begin{proposition}[Tahmasbi, Shahrasbi, and Gohari~\cite{Tahmasbi--Shahrasbi--Gohari2014b}]
\label{prop:dir_cycle}
If edge $e$ is critical for the side information graph $G$, then it lies on a directed cycle.
Thus, if the graph $G$ is critical, then it must be strongly connected.
\end{proposition}

However, belonging to a directed cycle is not a sufficient condition for an edge to be critical.
For the index coding problem shown in Fig. \ref{fig:3-message}, although the edge $2 \to 3$  lies on a directed cycle, it is not critical.

Side information subsets $A_1, \ldots, A_n$ of an index coding problem $G$ are said to be \emph{nondegraded} if for any $i \in A_j$, we have $A_i \not \subseteq A_j$.
In \cite{Arbabjolfaei--Kim2015}, nondegradedness is indicated as another necessary condition for criticality of an edge.

\begin{proposition}
\label{prop:degr_sideinfo}
If edge $i \to j$ is critical for the side information graph $G$, then $A_i \not \subseteq A_j$.
Thus, if the graph $G$ is critical, then side information subsets must be nondegraded.
\end{proposition}

\begin{figure}[h]
\vspace{-1pt}
\begin{center}
\small
\psfrag{1}[cb]{1}
\psfrag{2}[lb]{2}
\psfrag{3}[lt]{3}
\psfrag{4}[rt]{4}
\psfrag{5}[rb]{5}
\includegraphics[scale=0.4]{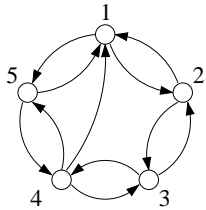}
\end{center}
\caption{A 5-node index coding problem. The edge $4 \to 1$ lies on a directed cycle and $A_4 \not \subseteq A_1$.
However, removing this edge does not affect the capacity region.
The capacity regions before and after removing this edge are achieved by the composite coding scheme \cite{Arbabjolfaei--Bandemer--Kim--Sasoglu--Wang2013}.}
\label{fig:ex1}
\vspace{-5pt}
\end{figure}

Satisfying the above two necessary conditions at the same time is still not a sufficient condition for an edge to be critical.
As an example, consider the side information graph shown in Fig.~\ref{fig:ex1}. 
The edge $4 \to 1$ satisfies the conditions in Propositions~\ref{prop:dir_cycle} and~\ref{prop:degr_sideinfo} at the same time, i.e., it lies on a directed cycle and $A_4 \not \subseteq A_1$. It can be shown, however, that it is not critical.

In \cite{Tahmasbi--Shahrasbi--Gohari2014}, Tahmasbi, Shahrasbi, and Gohari presented a simple sufficient condition for criticality of an edge.

\begin{proposition}
\label{prop:bidir}
Every bidirectional edge (either of a directed edge pair $i \to j$ and $j \to i$)
of the side information graph $G$ is critical; thus any side information graph consisting 
entirely of bidirectional edges
is critical.
\end{proposition}

The following critical graph structures are also identified in \cite{Tahmasbi--Shahrasbi--Gohari2014}.

\begin{proposition}
\label{prop:critical-struct}
\mbox{} \begin{enumerate}
\item Let $V = [1:n]$ and $E = \{(j+1,j): 1 \leq j \leq n-1\} \cup \{(1,n)\}$.
Construct a new graph $G'=(V',E')$, where $V' = V \cup \{n+1\}$ and $E' = E \cup \{(1,n+1),(n+1,i),(j,n+1),(n+1,k)\}$, for some $i,j$, and $k$ such that $1 \leq i < j \leq k \leq n$. Then $G'$ is critical.\\

\item Given a graph $G'$ that satisfies the condition in part 1, construct a new graph $G''=(V'',E'')$ by replacing any vertex $u \in V'$ by a complete graph (different vertices can be replaced by complete graphs of different sizes). Then $G^{''}$ is critical. 
More precisely, vertex $u$ is replaced with $n_u$ vertices $u^{(1)}, \ldots, u^{(n_u)}$ that form a complete graph.
There is a directed edge in $G''$ from node $u^{(i)}$, $i \in [1:n_u]$, to node $v^{(j)}$, $j \in [1:n_v]$, if and only if there exists a directed edge from $u$ to $v$ in $G'$.
\end{enumerate}
\end{proposition}

\begin{remark}
The conditions in Proposition \ref{prop:critical-struct} also imply criticality with respect to the symmetric capacity (in addition to criticality with respect to the capacity region).
\end{remark}

\begin{figure}[h]
\vspace{-5pt}
\begin{center}
\small
\psfrag{1}[cb]{1}
\psfrag{2}[lb]{2}
\psfrag{3}[lt]{3}
\psfrag{4}[rt]{4}
\psfrag{5}[rb]{5}
\includegraphics[scale=0.4]{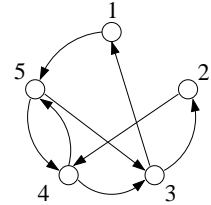}
\end{center}
\caption{A 5-node critical index coding problem which has both bidirectional and unidirectional edges and does not have the structure of Proposition \ref{prop:critical-struct}.}
\label{fig:5-node-uouc}
\vspace{-5pt}
\end{figure}

There are many critical graphs that are neither bidirectional nor in the form of Proposition \ref{prop:critical-struct} (see Fig. \ref{fig:5-node-uouc} for an example).

In this paper, we introduce the notion of unicycle for the side information graph and show that an edge is critical if it belongs to a unicycle.
This sufficient condition is more general than the existing sufficient conditions and indeed
both Propositions~3 and~4 are simple corollaries.
Unfortunately, even with this strengthening, the new sufficient condition is not necessary in general.
We clarify additional condition under which this sufficient condition becomes necessary, which
can be viewed as a partial converse.

Among all $n$-node index coding problems, critical graphs are only a fraction and the notion
of criticality has an immediate utility in reducing the number of instances that require 
analytical or numerical investigation.
Our sufficient and necessary conditions show that:
\begin{enumerate}
\item At most 411 out of 9,608 instances of 5-node index coding problems 
are critical.
\item Every edge belongs to a unicycle in 115 instances; thus there are at least 115 critical problems.
\end{enumerate}

The rest of the paper is organized as follows.
We first review some of the existing bounds on the capacity region in Section \ref{sec:bounds}.
The main result of the paper, the unicycle sufficient condition, is presented in Section \ref{sec:unicycle}.
In Section \ref{sec:converse}, a partial converse for the theorem of Section \ref{sec:unicycle} is provided which gives a new necessary condition for criticality of an edge.
In Section \ref{sec:circularclass}, we establish the capacity region of a class of index coding problems for which the unicycle condition fully characterizes the critical edges.
Throughout the paper,  $G|_S$  denotes the vertex induced subgraph of $G=(V,E)$ for $S \subseteq V$.

\section{Some Bounds on the Capacity Region}
\label{sec:bounds}

The following outer bound \cite{Arbabjolfaei--Bandemer--Kim--Sasoglu--Wang2013} is a special case of the polymatroidal outer bound \cite{Dougherty--Freiling--Zeger2011, Blasiak--Kleinberg--Lubetzky2011} and a slight generalization of the bound on the symmetric capacity by Bar-Yossef, Birk, Jayram, and Kol \cite{Bar-Yossef--Birk--Jayram--Kol2011} to the capacity region. 
\begin{proposition}[Maximal acyclic induced subgraph (MAIS) outer bound]
\label{prop:cfbound}
If rate tuple $(R_1, \ldots, R_n)$ is achievable for index coding problem $G$, then
\begin{align}
\label{eq:cfbound}
\sum_{j \in S} R_j \leq 1, \quad \forall S ~\text{s.t.}~ G|_S ~\text{is acyclic}. 
\end{align}
\end{proposition}

\medskip

Let $\Kc$ be the collection of all cliques of side information graph $G$.
The following proposition is a generalization of the fractional local clique covering bound on the symmetric capacity introduced by Shanmugam, Dimakis, and Langberg \cite{Shanmugam--Dimakis--Langberg2013b}.

\begin{proposition}[Fractional local clique covering inner bound]
\label{prop:flcc}
A rate tuple $(R_1, \ldots, R_n)$ is achievable for the index coding problem $(j \cond A_j), j \in [1:n]$, if there exists $\left( \rho_S \in [0,1], S \in \Kc \right)$ such that
\begin{equation}
\label{eq:flcc}
\begin{split}
\max_{j \in [1:n]}& \sum_{S \in \Kc: S \not \subseteq A_j} \rho_S \leq 1,\\
& \sum_{S \in \Kc: j \in S} \rho_S \ge R_j, \quad j \in [1:n].
\end{split}
\end{equation}
\end{proposition}

Tighter inner bounds can be found
in~\cite{Arbabjolfaei--Bandemer--Kim--Sasoglu--Wang2013, Arbabjolfaei--Kim2014}.
We only need the simpler inner bound of Proposition \ref{prop:flcc} 
for the purpose of this paper.

\bigskip

\section{The Unicycle Sufficient Condition}
\label{sec:unicycle}

We start with the definition of the \emph{unicycle}.
 
\begin{definition}
A graph $G=(V,E)$ is referred to as a \emph{unicycle} if the set of edges $E$ of the graph is a Hamiltonian cycle of $G$.
\end{definition}

\begin{figure}[h!]
\begin{center}
\psfrag{1}[cb]{1}
\psfrag{2}[lb]{4}
\psfrag{3}[lt]{3}
\psfrag{4}[rt]{2}
\includegraphics[scale=0.4]{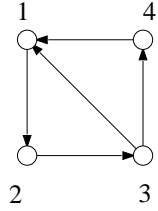}
\end{center}
\caption{A 4-node side information graph $G$. $G|_{\{1,2,3\}}$ is a unicycle, but $G$ is not a unicycle.}
\label{fig:ex2}
\end{figure}

Note that if the graph $G$ is a unicycle, then no proper subgraph can be a unicycle.
As an example, in Fig.~\ref{fig:ex2}, the vertex induced subgraph $G|_{\{1,2,3\}}$ is a unicycle, but $G$ itself is not a unicycle.

\begin{theorem}
\label{thm:critical-edge}
An edge $e$ of side information graph $G=(V,E)$ is critical if it belongs to a vertex induced subgraph of $G$ which is a unicycle.
\end{theorem}

\begin{IEEEproof} 
Let $e$ be an edge of $G|_S$, where $S \subseteq V$ and $G|_S$ is a unicycle.
Rate tuple $(R_1, \ldots, R_n)$ defined by 
\begin{align}
\label{eq:R}
R_i = \begin{cases} 
0, &  i \not \in S, \\
\frac{1}{|S|-1}, & i \in S,
\end{cases}
\end{align}
is achievable for index coding problem $G$ by a simple scalar linear code (since $G|_S$ has a Hamiltonian cycle).
Let $G'$ be the graph resulting from removing $e$ from $G$. 
The vertex-induced subgraph $G'|_S$ is acyclic (since the Hamiltonian cycle of $G|_S$ is broken and by definition there is no other cycle).
Therefore, due to the MAIS outer bound, any rate tuple $(R'_1, \ldots, R'_n)$ in the capacity region of $G'$ satisfies
\begin{align}
\label{eq:cf}
\sum_{i \in S} R'_i \leq 1.
\end{align}
The rate tuple defined in \eqref{eq:R} does not satisfy \eqref{eq:cf} and thus is not achievable for index coding problem $G'$.
This means that removing edge $e$ from $G$ strictly shrinks the capacity region and hence $e$ is a critical edge of $G$.
\end{IEEEproof}

In the rest of the paper, when edge $e$ belongs to a vertex induced subgraph which is a unicycle, we briefly say edge $e$ belongs to a unicycle.
The following corollary of Theorem \ref{thm:critical-edge} establishes a sufficient condition for a graph to be critical.

\begin{corollary}
\label{cor:unicycle-union}
If every edge of the side information graph $G$ belongs to a unicycle, then $G$ is critical.
\end{corollary}

\begin{remark}
Corollary \ref{cor:unicycle-union} includes the critical structures in Proposition \ref{prop:critical-struct} as special cases.
\end{remark}

\begin{figure}[h!]
\begin{center}
\psfrag{1}[cb]{1}
\psfrag{2}[lb]{4}
\psfrag{3}[lt]{3}
\psfrag{4}[rt]{2}
\includegraphics[scale=0.4]{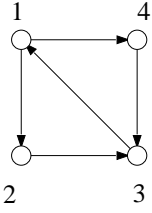}
\end{center}
\caption{A 4-node side information graph. $G|_{\{1,2,3\}}$ and $G|_{\{1,3,4\}}$ are both unicycles.}
\label{fig:ex3}
\end{figure}

In the graph shown in Fig.~\ref{fig:ex3}, every edge belongs to a unicycle and hence it is critical.

\begin{figure}[h]
\begin{center}
\small
\psfrag{1}[cb]{1}
\psfrag{2}[lb]{2}
\psfrag{3}[lt]{3}
\psfrag{4}[rt]{4}
\psfrag{5}[rb]{5}
\includegraphics[scale=0.4]{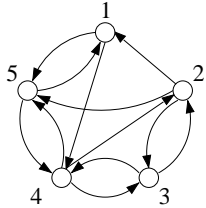}
\end{center}
\caption{A critical 5-node index coding problem.
Although the edge $2 \to 5$ does not belong to any unicycle, it is critical.
The capacity regions before and after removing this edge are achieved by composite coding \cite{Arbabjolfaei--Bandemer--Kim--Sasoglu--Wang2013}.}
\label{fig:noncritical}
\end{figure}

The converse to Theorem \ref{thm:critical-edge} does not hold in general, 
i.e., there exist side information graphs with critical edges that do not belong to any unicycle.
One such example is shown in Fig.~ \ref{fig:noncritical}.

\section{A Partial Converse}
\label{sec:converse}

Throughout this subsection, we assume that $e$ is an edge of side information graph $G=(V,E)$, and denote the graph resulting from removing $e$ from $G$ by $G_e$.
Let $\Cr$ and $\Cr_e$ be the capacity regions of $G$ and $G_e$ respectively.
The following lemma shows that the notion of unicycle captures ``criticality'' not with respect
to the capacity region, but with respect to the MAIS outer bound.

\begin{lemma}
\label{lem:unicycle-cf}
Edge $e$ belongs to a unicycle if and only if the MAIS bound on $\Cr_e$ is a proper subset of the MAIS bound on $\Cr$.
\end{lemma}

\begin{IEEEproof}
\emph{Sufficiency.}
If the MAIS bound on $\Cr_e$ is a proper subset of the MAIS bound on $\Cr$, there exists a subset $S \subseteq V$ such that $G|_S$ contains a cycle and $G_e|_S$ is acyclic.
Let $S_\text{min}$ be a minimal such subset.
Then, $G|_{S_\text{min}}$ is a unicycle that has $e$ as one of its edges.

\emph{Necessity.} 
Let $G|_S$, $S \subseteq V$ be a unicycle that has $e$ as an edge.
By definition of a unicycle, the graph $G_e|_S$ is acyclic.
Therefore, inequality  
\begin{align*}
\sum_{j \in S} R_j \leq 1,
\end{align*} 
is an inequality of the MAIS bound on $\Cr_e$, but is not an inequality of the MAIS bound on $\Cr$, which completes the proof of the lemma.
\end{IEEEproof}

Next, we use Lemma \ref{lem:unicycle-cf} to establish the following partial converse to Theorem \ref{thm:critical-edge}.

\begin{theorem} 
\label{thm:converse}
If edge $e$ is critical for side information graph $G$, then 
\begin{enumerate}
\item it belongs to a unicycle, 
or
\item it does not belong to a unicycle and the MAIS bound is not tight for $G_e$.
\end{enumerate}
\end{theorem}

\begin{IEEEproof}
It suffices to show that if $e$ is a critical edge of $G$ that does not belong to any unicycle, then the MAIS bound is not tight for index coding problem $G_e$.
Since $e$ is a critical edge for $G$, we have $\Cr_e \subsetneq \Cr$.
Assume by contradiction that MAIS bound is tight for $G_e$.
This makes the MAIS bound on $\Cr_e$ to be a proper subset of the MAIS bound on $\Cr$.
Thus, by Lemma \ref{lem:unicycle-cf}, $e$ belongs to a unicycle, which contradicts the assumption.
\end{IEEEproof}

\section{A Class of Index Coding Problems}
\label{sec:circularclass}

In this section, we consider the class of index coding problems with side information subsets $A_1, \ldots, A_n$ satisfying
\begin{equation}
\label{eq:cycle}
A_j \subseteq \{j-1,j+1\}, \quad j \in [1:n].
\end{equation}

The following proposition characterizes the capacity region of the index coding problems in this class for which at least one $A_j$ is a proper subset of $\{j-1,j+1\}$.

\begin{proposition}
\label{prop:cf-tight}
The MAIS outer bound is tight for index coding problem $G$ if \eqref{eq:cycle} is satisfied and
\begin{equation}
\label{eq:proper}
A_j \subsetneq \{j-1,j+1\}, \quad \text{for some}~ j \in [1:n].
\end{equation}
\end{proposition}
The proof of the proposition is presented in the Appendix.
For the class of index coding problems satisfying \eqref{eq:cycle}, the converse to Theorem \ref{thm:critical-edge} holds.

\begin{proposition}
\label{prop:circular}
Given an index coding problem $G=(V,E)$ satisfying \eqref{eq:cycle}, an edge $e \in E$ is critical if and only if it belongs to a unicycle.
\end{proposition}

\begin{IEEEproof}
The sufficiency follows from Theorem \ref{thm:critical-edge}.
For necessity, by Proposition \ref{prop:cf-tight}, for any edge $e \in E$, the MAIS bound is tight for $G_e$.
Thus, Theorem \ref{thm:converse} implies that every critical edge must belong to a unicycle.
\end{IEEEproof}

\begin{figure}[h]
\smallskip
\captionsetup[subfigure]{aboveskip=1em,belowskip=0.5em,justification=centering}
\begin{subfigure}[c]{0.48\linewidth}
\centering
\small
\psfrag{1}[cb]{1}
\psfrag{2}[lb]{2}
\psfrag{3}[lt]{3}
\psfrag{4}[rt]{4}
\psfrag{5}[rb]{5}
\includegraphics[scale=0.4]{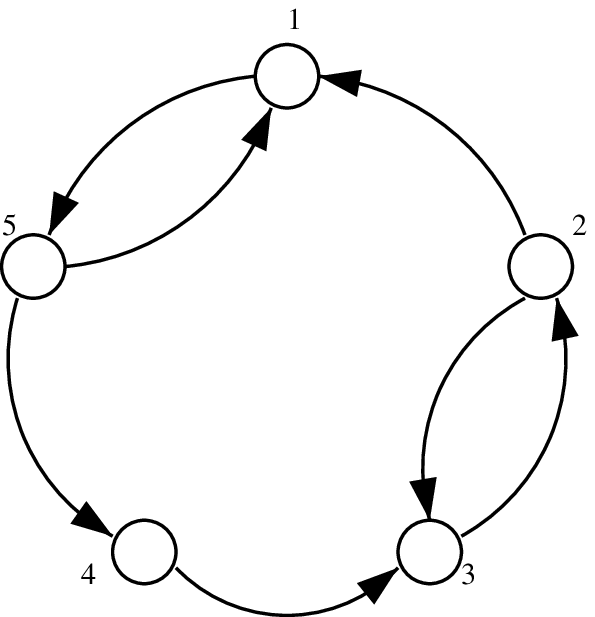}
\caption{}
\end{subfigure}
\begin{subfigure}[c]{0.48\linewidth}
\small
\psfrag{1}[cb]{1}
\psfrag{2}[lb]{2}
\psfrag{3}[lt]{3}
\psfrag{4}[rt]{4}
\psfrag{5}[rb]{5}
\qquad~\includegraphics[scale=0.4]{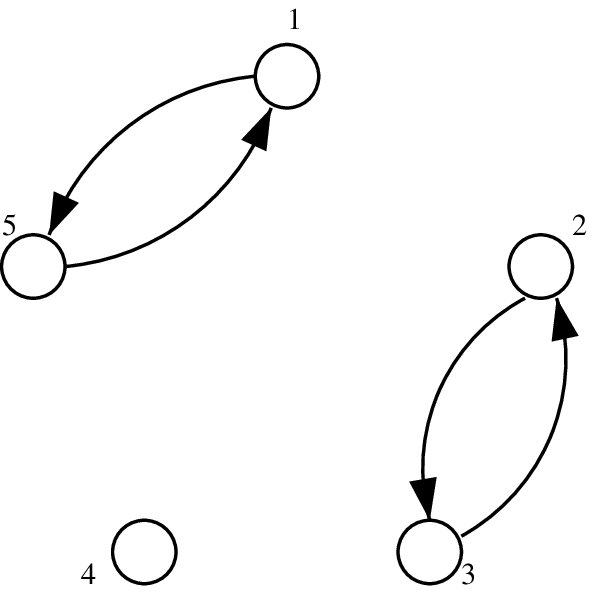}
\caption{}
\end{subfigure}
\caption{Two 5-node index coding problems with the same capacity region.}
\label{fig:same-region}
\end{figure}

In the side information graph shown in Fig. \ref{fig:same-region} (a), edges $5 \to 4 $, $4 \to 3 $, and $2 \to 1 $ do not belong to any unicycle.
Hence, the two side information graphs shown in Fig. \ref{fig:same-region} have the same capacity region.

\section{Acknowledgments}
This work was supported by the Korean MSIP under 
IITP Grant B0132-15-1005 (ETRI) and by the National Science Foundation under Grant CCF-1320895.


\appendix[Proof of Proposition \ref{prop:cf-tight}]
\label{sec:proof}

To prove the proposition, we will show that for this class of index coding problems, the fractional local clique covering inner bound matches the MAIS outer bound.
To do this, for any rate tuple $(R_1, \ldots, R_n)$ that satisfies the MAIS bound in \eqref{eq:cfbound},  we explicitly construct $\left( \rho_S \in [0,1], S \in \Kc \right)$ that meets the conditions in Proposition \ref{prop:flcc}.

For side information graphs satisfying \eqref{eq:cycle}, there are only two possible types of cycles,  namely, Hamiltonian cycles and length-two cycles.
Based on this, we consider the following three cases.

\emph{Case 1.} If $G$ is acyclic, \eqref{eq:cfbound} implies
$
\sum_{j=1}^n R_j \leq 1.
$
In this case, the size of the largest clique is one and $\left(
\rho_{\{j\}} = R_j,~ j \in [1:n] \right)$ 
satisfies \eqref{eq:flcc}.

\emph{Case 2.} If $G$ has exactly one Hamiltonian cycle and has no cycle of length two, then \eqref{eq:cfbound} implies
$
\sum_{j \in [1:n], j \neq i} R_j \leq 1, \quad i \in [1:n].
$
Similar to case 1, the size of the largest clique is one and $\left(
\rho_{\{j\}} = R_j, j \in [1:n] \right)$
satisfies \eqref{eq:flcc}.

\emph{Case 3.} If $G$ has at least one cycle of length two, then there will be some cliques of size two in addition to the cliques of size one.
A set of nodes $\{i, i+1, \ldots, i+k\}$ is said to form a chain of length $k$ if 
\begin{align*}
A_j = \begin{cases}
\{i+1\}, & j = i, \\
\{j-1,j+1\}, & i+1 \leq j \leq i+k-1, \\
\{i+k-1\}, & j = i+k.
\end{cases}
\end{align*}
We first assign $\rho_S$ to cliques of size one as follows:
\begin{align*}
\rho_{\{j\}} = \begin{cases}
R_j, & \text{if node $j$ does not belong to any chain}, \\
0, & \text{otherwise}.
\end{cases}
\end{align*}
Any clique of size two belongs to a chain.
For each chain, Algorithm 1 is used to assign $\rho_S$ to the cliques of size two.
The algorithm ensures that no two consecutive $R_j$'s of a chain appear in the total sum.
Therefore,
\[
\sum_{S \in \Kc} \rho_S = \sum_{j \in V'} R_j,
\]
for some $V' \subseteq V$ such that $G|_{V'}$ is acyclic.
Hence, $(\rho_S \in [0,1], S \in \Kc)$ as constructed above satisfies \eqref{eq:flcc}.
This completes the proof of the proposition.
\begin{algorithm}
\caption{}

{\bf input:} A set of nodes $\{i, i+1, \ldots, i+k\}$ that form a chain of length $k$ \\
{ \bf output:} $(\rho_{\{j,j+1\}}, j \in [i:i+k-1])$ \\

\noindent Step 1) If $k=1$, set $\rho_{\{i,i+1\}} = \max\{R_i,R_{i+1}\}$ and we are done. Otherwise, go to step 2. \\

\noindent Step 2) If $k=2$,
\begin{itemize}
\item if $R_{i+1} \leq R_i + R_{i+2}$, set $\rho_{\{i,i+1\}} = R_i$, and $\rho_{\{i+1,i+2\}} = R_{i+1}$,
\item if $R_{i+1} > R_i + R_{i+2}$, set $\rho_{\{i,i+1\}} = R_i$ and $\rho_{\{i+1,i+2\}} = R_{i+1} - R_i$,
\end{itemize}
and we are done. Otherwise, go to step 3. \\

\noindent Step 3) If $k \geq 3$,
\begin{itemize}
\item if $R_i > R_{i+1}$, set $\rho_{\{i,i+1\}} = R_i$, and $\rho_{\{i+1,i+2\}} = 0$ and repeat the algorithm for the chain $\{i+2, \ldots, i+k\}$,
\item if $R_{i+k} > R_{i+k-1}$, set $\rho_{\{i+k-1,i+k\}} = R_{i+k}$, and $\rho_{\{i+k-2,i+k-1\}} = 0$ and repeat the algorithm for the chain $\{i, \ldots, i+k-2\}$,
\item if $R_i < R_{i+1}$ and $R_{i+k} < R_{i+k-1}$, set $\rho_{\{i,i+1\}} = R_i$ and $\rho_{\{i+k-1,i+k\}} = R_{i+k}$, and repeat the algorithm for the chain $\{i+1, \ldots, i+k-1\}$, with the following new values: $R_{i+1} \leftarrow R_{i+1}-R_i$ and $R_{i+k-1} \leftarrow R_{i+k-1} - R_{i+k}$.
\end{itemize}

\end{algorithm}

\bibliographystyle{IEEEtran}
\bibliography{nit} 

\end{document}